\documentclass[intlimits,twoside,a4paper]{article}

\usepackage[cp1251]{inputenc}

\usepackage[eqsecnum]{cmpj3}

\usepackage{subfigure}
\usepackage{bm}

\issue{2019}{22}{1}{13001}
\doinumber{10.5488/CMP.22.13001}

\title[On the finite-size effects in two segregated Bose-Einstein condensates restricted by a hard wall]%
{On the finite-size effects in two segregated Bose-Einstein condensates restricted by a hard wall}
\author[H.V. Quyet, N.V. Thu, D.T. Tam, T.H. Phat]{H.V. Quyet\refaddr{label1}, N.V. Thu\refaddr{label1}, D.T. Tam\refaddr{label2}, T.H. Phat\refaddr{label3}}
\addresses{
\addr{label1}Department of Physics, Hanoi Pedagogical University 2, Hanoi, Vietnam 
\addr{label2} Tay Bac University, Son La, Vietnam 
\addr{label3} Vietnam Atomic Energy Commission, 59 Ly Thuong Kiet, Hanoi, Vietnam 
}

\date{Received November 21, 2018, in final form February 2, 2019}

\begin{document}

\maketitle

\begin{abstract}
The finite-size effects in two segregated Bose-Einstein condensates (BECs) restricted by a hard wall is studied by means of the Gross-Pitaevskii equations in the double-parabola approximation (DPA). Starting from the consistency between the boundary conditions (BCs) imposed on condensates in  confined geometry and in the full space, we find all possible BCs together with the corresponding  condensate profiles and interface tensions. We discover two finite-size effects: a) The ground state derived from the Neumann BC is stable whereas the ground states derived from the Robin and Dirichlet BCs are unstable. b) Thereby, there equally manifest two possible wetting phase transitions originating from two unstable states. However, the one associated with the Robin BC is more favourable because it corresponds to a smaller interface tension.  
\pacs 03.75.Hh, 05.30.Jp, 68.03.Cd, 68.08.Bc
\end{abstract}

\section{Introduction}\label{1}
The physics of binary Bose-Einstein condensate mixture in infinite space has explored many interesting instabilities. It was proved theoretically that the well-known instabilities in classical fluid also take place in quantum fluid, including the Rayleigh-Taylor instability \cite{c1, c2}, the Kelvin-Helmholtz instability \cite{c3, c4}, the Richtmyer-Meshkov instability \cite{c5}, and the Rayleigh-Plateau instability \cite{c6} for immiscible case. For the miscible case, the counter-superflow instability was discovered in \cite{c7, c8, c9} and the dynamical instability was recently predicted to occur in both  single-component \cite{c10} and two-component Bose-Einstein condensates (BECs) \cite{c11}. This instability is associated with the   superfluidity breaking of BECs and, therefore, it is purely a quantum effect. In the semi-infinite space where the system of two segregated BECs is restricted by a hard wall, its ground state is fully determined by the boundary conditions (BCs) at hard wall. In  \cite{c12n, c13n}, with the Dirichlet BCs imposed for both condensates, the authors indicated impeccably that the system of interest can undergo a wetting phase transition. In the present paper we explore the unstable states  associated with the BCs at a hard wall, and then we  proceed to the study of wetting phase transitions. Here, we prove that the wetting phase transition associated with the Robin BC at hard wall is more favourable.

\begin{figure}[!t]
	\begin{center}
		\includegraphics[scale=0.22]{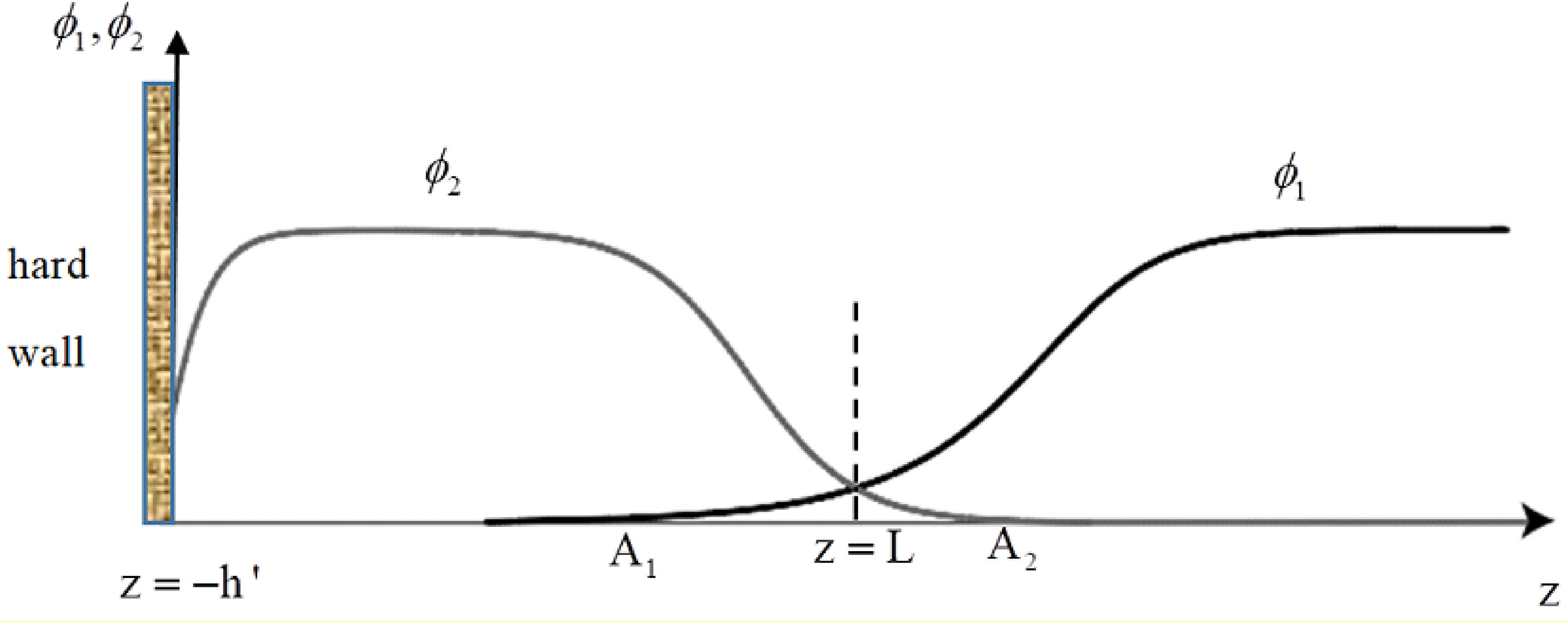}
		\caption{(Colour online) The hard wall is located at $z = -  h'$, the interface at $z = L$ and the component 1 (2) occupies the region $z > L$ $(z < L )$. The curve $j$ represents the condensate profile $j$ and the interval $L{A_j}$ is the penetration depth of the condensate $j$. }\label{f1}
	\end{center}
\end{figure}

To this end, we  make use of the double-parabola approximation (DPA) method which was proposed in  \cite{c12, c13} and was developed in \cite{c14, c15, c16}. The boundary problem we deal with is schematically given in figure~\ref{f1} where the hard wall (optical wall) is placed at $z = - h'$, the interface at $z = L$ and the curve $j$ $(j = 1, 2)$ represent the condensate profile ${\phi _j}( z )$ which will be introduced in the next section. The interval $L{A_j}$ is  the so-called penetration depth of condensate $j$ \cite{c12} inside the condensate $j' \ne j.$ Its expression reads $$L{A_j} = {\xi _j}/\sqrt {K - 1} $$ with $${\xi _j} = \hbar /\sqrt {2{m_j}{\mu _j}} $$ being the healing length of condensate $j$. The parameters $K$, ${m_j}$ and ${\mu _j}$ will be cleared up in what follows. Therefore, the interval ${A_1}{A_2}$ is the coexistence region of two condensates.

Following \cite{c17, c18}, $H$  can be written in  the form
\begin{equation}\label{HMT1}
H = \int\limits_V {\rd V{H_\text{GP}}}  + \int\limits_W {\rd S{H_\text{W}}}\,, 
\end{equation}     
here, the Gross-Pitaevskii (GP) Hamiltonian ${H_\text{GP}}$ in the bulk and the wall Hamiltonian ${H_\text{W}}$ are given as follows:\\
\indent -- The GP Hamiltonian
\begin{subequations}\label{HMT2}
	\begin{align}
	&\hspace{-0.5cm}{H_\text{GP}} = \sum\limits_{j = 1}^2 {{{\psi_j^*}\left( {-\frac{{{\hbar ^2}}}{{2{m_j}}}{\nabla ^2}} \right)} {\psi_j} + V\left( {{\psi _1},{\psi _2}} \right)} ,\qquad\qquad\\
	&\hspace{-3.7cm}\text{where}\notag\\
	&\hspace{-0.5cm}V\left( {{\psi _1},{\psi _2}} \right) = \sum\limits_{j = 1}^2\left(  { - {\mu _j}\psi_j^*{\psi _j} + \frac{{{g_{jj}}}}{2}{{\left| {{\psi _j}} \right|}^4}}\right)   + {g_{12}}{\left| {{\psi _1}} \right|^2}{\left| {{\psi _2}} \right|^2},
	\label{the1}
	\end{align}
\end{subequations}
here, ${\psi _j} = {\psi _j}(\vec x)$, ${m_j}$ and ${\mu _j}$ $(j = 1,2)$ are respectively the wave function, the atomic mass and the chemical potential of each species $j$, with $\vec x = (\vec a, z)$, $\vec a = (x,y)$ to be denoted in what follows.
For inhomogeneous state, in which two condensates are separated by interface, we have $K=\frac{g_{12}}{\sqrt{{g_{11}}{g_{22}}}}  > 1$ and hereafter we will consider the case of all coupling constants being positive. \\
\indent --	The wall Hamiltonian. Usually, the wall Hamiltonian is chosen in the phenomenological form \cite{c17, c18}
\begin{align}\label{HMT3}
{H_\text{W}} = \sum\limits_{j = 1}^2 {\frac{{{\hbar ^2}}}{{2{m_j}{\lambda _j}}}} \psi _{\text{W}j}^ * {\psi _{\text{W}j}}\,,
\end{align}
where ${\psi _{\text{W}j}}$ $(j = 1, 2)$ is the wall field induced by ${\psi _j}$ on the wall and  ${\lambda _j}$ is its extrapolation length, introduced by de Gennes \cite{c19}. Next, let us pay special attention to the wall fields when the wall is flat. For the first condensate, figure~\ref{f1} indicates that the Dirichlet BC
$${\psi _1}( {\vec a,z =  - h'} ) = 0$$
is justified when $h' + L \geqslant \frac{\xi _1}{\sqrt {K - 1}} $ or $K \geqslant \frac{{\xi _1^2}}{{{{\left( {h' + L} \right)}^2}}} + 1$.

Moreover, the Dirichlet BC is also consistent with the condition in the infinite space ${\psi _1}( {\vec a, - \infty } ) = 0$ \cite{c14, c16}. As to the second condensate, we have two options:\\
either $${\psi _2}\left( {\vec a,z =  - h'} \right) = 0,$$
or $${\psi _2}\left( {\vec a,z =  - h'} \right) \ne 0.$$
It is easily seen that the consistency between the BCs in the semi-infinite space and in the infinite space eliminates the first option. Indeed, if $h' \to \infty $ one obtains $${\lim _{h' \to \infty }}{\psi _2}\left( {\vec a,z =  - h'} \right) = {\psi _2}\left( {\vec a, - \infty } \right) = 0,$$ 
which contradicts the BC for this condensate in the infinite space \cite{c12, c20}, ${\psi _2}\left( {\vec a, - \infty } \right) \ne 0.$ 

Hence, in reality the wall Hamiltonian (\ref{HMT3}) is simplified to
\begin{align}\label{HMT4}
{H_\text{W}} = \frac{{{\hbar ^2}}}{{2{m_2}{\lambda _2}}}\psi _2^ * \left( {\vec a,z =  - h'} \right){\psi _2}\left( {\vec a,z =  - h'} \right)
\end{align}
for the flat wall.

The fact that the equilibrium values of the fields minimizes the total Hamiltonian given in (\ref{HMT1}), (\ref{HMT2}) and (\ref{HMT4}) leads to the time-independent GP equations in the bulk 
\begin{subequations}\label{GP1}
	\begin{align}
	&\bigg( -\frac{\hbar^2}{2m_1}\Delta-\mu_1+ g_{11}| \psi_1|^2+g_{12}| \psi_{2} |^2 \bigg)\psi_1=0,\\
	&\bigg( -\frac{\hbar^2}{2m_2}\Delta-\mu_2+ g_{22}| \psi_2|^2+g_{12}| \psi_{1} |^2 \bigg)\psi_2=0,
	\end{align}
\end{subequations}
together with the BCs at the wall 
$$\vec n\nabla {\psi_{\text{W}j}} = \frac{1}{{{\lambda _j}}}{\psi _{\text{W}j}}\,,$$
where $\vec n$ is the unit vector normal to the wall and pointing inside the system. For a flat wall, the foregoing condition turns out to be the Robin BC  
\begin{align}\label{BC1}
{\left( {\frac{{\partial {\psi _2}\left( {\vec a,z} \right)}}{{\partial z}}} \right)_{z =  - h'}} = \frac{1}{{{\lambda _2}}}{\psi _2}\left( {\vec a,z =  - h'} \right).
\end{align}
It is easily seen that (\ref{BC1}) is reduced to the Neumann BC                                             \begin{align}\label{BC2}
{\left( {\frac{{\partial {\psi _2}\left( {\vec a,z} \right)}}{{\partial z}}} \right)_{z =  - h'}} = 0,
\end{align}	
when ${\lambda _2}$ tends to infinity  and to the Dirichlet BC
\begin{align}
{\psi _2}\left( {\vec a,z =  - h'} \right) = 0,
\end{align}
when ${\lambda _2} \to 0$.

The present paper is organized as follows. In section~\ref{2},  employing the DPA, we calculate the analytical solutions of the GP equations obeying the Robin BC at a hard wall. Section~\ref{3} is devoted to the determination of the interface tension in the grand canonical ensemble (GCE) and the wetting phase transition. The conclusion is given in section~\ref{4}. 

\section{Ground state in DPA }\label{2}
For two segregated BECs, the translation along the $Oz$ axis is spontaneously broken by the presence of the domain wall-the interface, and one usually focuses on the condensate profiles that only depend on $z$ and for simplicity we do not change their symbols $${\psi _j} = {\psi _j}( z),$$ which are assumed to be real without loss of generality. In terms of ${\psi _j}\left( z \right)$, the GP Hamiltonian (\ref{HMT2}) becomes
\begin{subequations}\label{HMT5}
	\begin{align}
	& H_\text{GP} = \sum\limits_{j=1}^{2}\psi_j\bigg(-\dfrac{\hbar^{2}}{2m_{j}} \frac{\rd^2}{\rd z^2}\bigg)\psi_j + V(\psi_1,\psi_2) \label{HMT03}\\
	&\hspace{-4.1cm} \text{ with } \notag\\
	&V(\psi_1,\psi_2)=\sum\limits_{j=1}^{2}\bigg( -\mu_j \psi_j^2+\frac{g_{jj}}{2}\psi _j^4 \bigg)+g_{12}\psi_1^2\psi_2^2. \label{the2}
	\end{align}
\end{subequations}
It is convenient to write all quantities in the dimensionless forms by introducing new quantities $\varrho  = z/{\xi _1}$, $h = h'/{\xi _1}$, $\ell = L/{\xi _1}$ and ${\phi _j} = {\psi _j}/\sqrt {{n_j}},$ where the densities ${n_j}$ $(j = 1,2)$ were determined in \cite{c14}. In the dimensionless forms, the GP  Hamiltonian (\ref{HMT5}) is rewritten as follows:
\begin{subequations}\label{HMT6}
	\begin{align}
	& \tilde H\left( {{\phi _1},{\phi _2}} \right) = \sum\limits_{j = 1}^2 {{\phi _j}} \left( { - \frac{{\xi _j^2}}{{{\xi _1}}}\frac{{{\rd^2}}}{{\rd{\varrho ^2}}}} \right){\phi _j} + \tilde V\left( {{\phi _1},{\phi _2}} \right),\label{HMT6a}\\
	&\hspace{-4.2cm} \text{ in which } \notag\\
	&\tilde V\left( {{\phi _1},{\phi _2}} \right) = \sum\limits_{j = 1}^2 {\left( { - \phi _j^2 + \frac{{\phi _j^4}}{2}} \right)}  + K\phi _1^2\phi _2^2\label{the3}.
	\end{align}
\end{subequations}
In the dimensionless form,  equations~(\ref{GP1}) read
\begin{subequations}\label{GP2}
	\begin{align}
	& -\frac{\rd^2\phi_1}{\rd\varrho^2}-{\phi_1}+\phi _1^3+K\phi _2^2\phi_1=0, \\
	& -\xi^2\frac{\rd^2\phi_2}{\rd\varrho^2}-\phi_2+\phi _2^3+K\phi _1^2\phi_2=0,
\end{align}
\end{subequations}
here, $\xi  = {\xi _2}/{\xi _1}$.

Correspondingly, the wall Hamiltonian (\ref{HMT4}) is rewritten in the form
\begin{align}\label{HMT7}
{H_\text{W}} &= \frac{{{\mu _2}\xi _2^2}}{{{\lambda _2}}}{n_{2}}\phi _2^2\left( { - h} \right).
\end{align}
Now, let us present in detail the formulation of the DPA \cite{c14}. The DPA method consists of three constituents:\\
\indent a) We temporarily introduce $\varrho  = \ell$  the interface location which separates the system into two regions as plotted in figure~\ref{f1}.\\
\indent b) The  foregoing potential (\ref{the3}) is linearized by the DPA potential ${V_\text{DPA}}( {{\phi _1},{\phi _2}})$ in appropriate half spaces separated by the interface
\begin{eqnarray}\label{the4}
\tilde V_\text{DPA}(\phi_j,\phi_2 )= 2(\phi_{j}-1)^2+\beta^{2} \phi_{j'}^2-\frac{1}{2}\,, \qquad
\left\{ 
\begin{aligned}
\hspace{-0.0cm}\varrho > \ell,\ (j, j') = (1,2),\\ 
\hspace{-0.0cm}\varrho < \ell,\ (j, j') = (2,1),\\ 
\end{aligned} 
\right.
\end{eqnarray}
where $ \beta=\sqrt{K-1}$.\\
\indent c) The smooth variation and the continuity of  both condensate profiles when passing the interface are assumed. Mathematically, this assumption is expressed by
\begin{align}\label{itf1}
&{\left( {\frac{{\rd{\phi _j}\left( \varrho  \right)}}{{\rd\varrho }}} \right)_{\varrho  = \ell - 0}} = {\left( {\frac{{\rd{\phi _j}\left( \varrho  \right)}}{{\rd\varrho }}} \right)_{\rho  = \ell + 0}},\\
&{\phi _j}\left( {\varrho  = \ell - 0} \right) = {\phi _j}\left( {\varrho  = \ell + 0} \right).
\end{align}     
It is important to note that the introduction of the interface position $\ell$ is merely a mathematical tactics based on which one sets up the DPA potential. Indeed, the interface is defined as the intersection of two condensates
\begin{align}\label{itf2}
{\phi _1}\left( {\varrho  = \ell;\ell,K,\xi } \right) = {\phi _2}\left( {\varrho  = \ell;\ell,K,\xi } \right),
\end{align}
which indicates that $\ell$ is a function of $K$ and $\xi $
\begin{align}\label{itf3}
\ell = f\left( {K,\xi } \right).
\end{align}    
Therefore, in reality, both condensates depend only upon $\varrho$ and the model parameters.

Taking into account  (\ref{HMT6a}), (\ref{HMT7}) and (\ref{the4}), we arrive at the total Hamiltonian in the DPA, ${H_\text{DPA}}$. Minimizing it, we are led to the GP equations in DPA together with the BCs, namely: \\
\indent -- GP equations. \\
Equations in the right-hand side of the interface
\begin{subequations}\label{GP3}
	\begin{align}
	&-\frac{\rd^2\phi_1}{\rd\varrho^2}+2(\phi_1-1)=0,\label{gp3a}\\
	& -\xi^2\frac{\rd^2\phi_2}{\rd\varrho^2}+\beta^2 \phi_2=0.\label{gp3b}\\
	&\hspace{-5.5cm}\text{Equations in the left-hand side of the interface  }\notag\\
	&  -\frac{\rd^2\phi_1}{\rd\varrho^2}+\beta^2\phi_1=0, \label{gp3c}\\
	& -\xi^2\frac{\rd^2\phi_2}{\rd\varrho^2}+2(\phi_2-1)=0.\label{gp3d}
	\end{align}
\end{subequations}
\indent --	The BCs that both condensate profiles fulfil are:\\
 The Dirichlet BC at hard wall for the first condensate
\begin{align}\label{BC4}
{\phi _1}\left( -h \right) = 0.
\end{align}
The Robin BC at hard wall for the second condensate
\begin{align}\label{BC5}
{\left( {\frac{{\rd{\phi _2}\left( \varrho  \right)}}{{\rd\varrho }}} \right)_{\varrho  =  - h}} = c{\phi _2}\left( { - h} \right) 
\end{align}
with $c = \frac{{{\xi _1}}}{{{\lambda _2}}}.$
 
Besides these, we still have the conditions for both condensates at infinity
\begin{align}\label{BC6}
{\phi _1}\left( { + \infty } \right) = 1, \qquad {\phi _2}\left( { + \infty } \right) = 0.
\end{align} 
The analytical expressions of the solutions of equations~(\ref{GP3}) which obey the conditions (\ref{BC4}), (\ref{BC5}) and (\ref{BC6}) are in order: \\
\indent -- In the RHS of the interface
\begin{subequations}\label{sl1}
	\begin{align}
	\phi_{1}(\varrho)&=1 + {A_1}{{\rm{e}}^{ - \sqrt 2  \varrho }},\label{solu1d}\\
	\phi_{2}(\varrho)&={A_2}{{\rm{e}}^{ - \frac{{\beta \varrho }}{\xi }}}.\label{solu2d}
	\end{align}
\end{subequations} 
\indent -- In the LHS of the interface
\begin{subequations}\label{sl2}
	\begin{align}
	\phi_{1}(\varrho)=&{B_1}{{\rm{e}}^{ - \beta \left( {2h + \varrho } \right)}}\left[ { - 1 + {{\rm{e}}^{2\beta \left( {h + \varrho } \right)}}} \right],\label{solu1a}\\
	\phi_{2}(\varrho)=&1+ {B_2}{{\rm{e}}^{\frac{{\sqrt 2 \varrho }}{\xi }}} + \frac{1}{{\sqrt 2  + c\xi }}{{\rm{e}}^{ - \frac{{\sqrt 2 \left( {2h + \varrho } \right)}}{\xi }}}\left[ {\sqrt 2 {B_2} - c\left( {{B_2} + {{\rm{e}}^{\frac{{\sqrt 2 h}}{\xi }}}} \right)\xi } \right].\label{solu2a}
	\end{align}
\end{subequations}
Substituting (\ref{sl1}), (\ref{sl2}) into (\ref{itf1}), (\ref{itf2}), we find  the analytical expressions of ${A_j}$, ${B_j}$ $(j =1,2)$ which  are given in appendix~\ref{A}.

\begin{figure}[!b] 
    \centering 
        \subfigure[$c = 1/\sqrt 2, h = 0,\xi  = 0.5, K = 2.5.$]{ 
         \includegraphics[width=5.9cm]{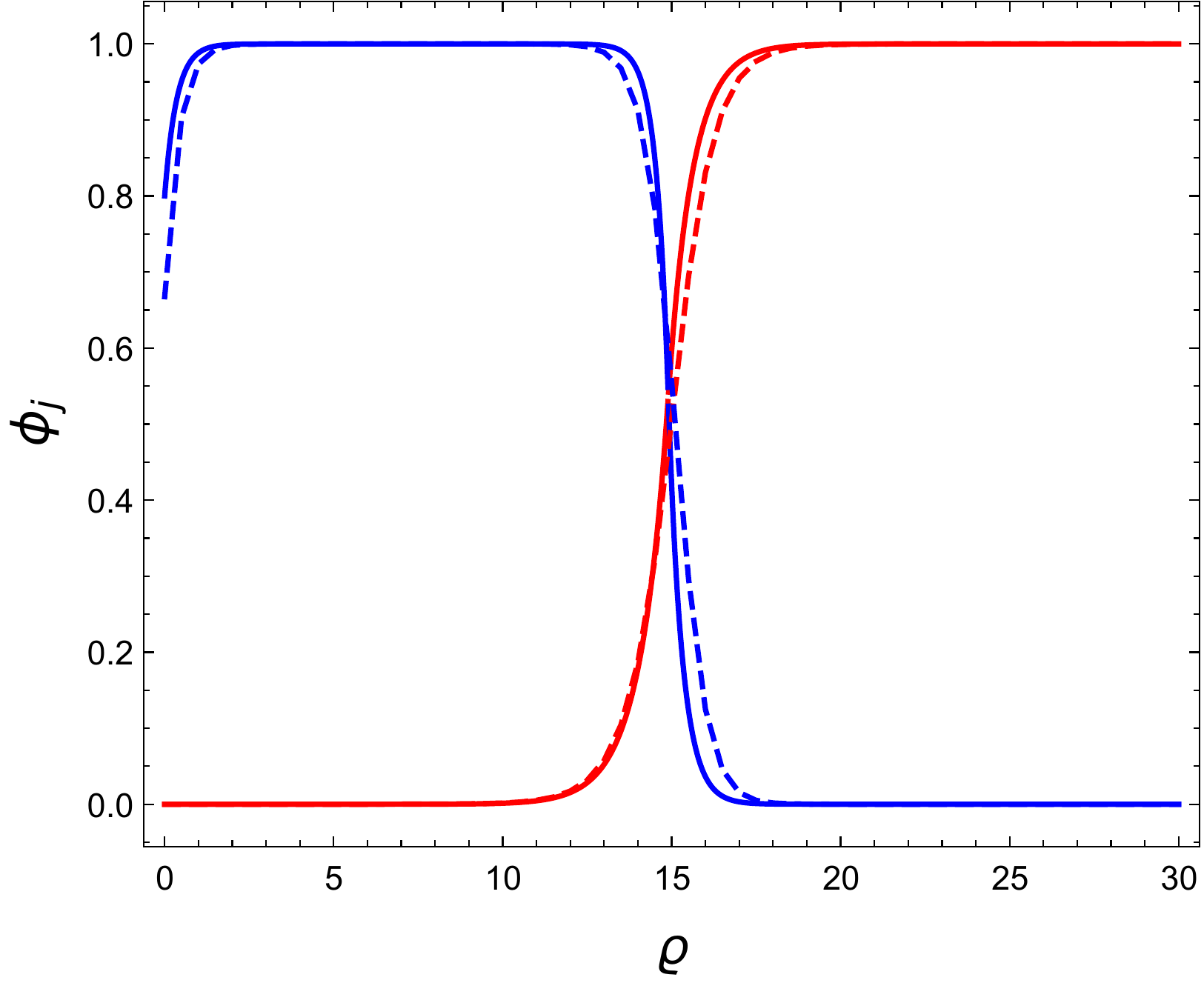} 
         \label{fig:sub-slotted-hopping}} \qquad
        \subfigure[$c = 1/\sqrt 2, h = 0,\xi  = 1.0, K = 3.$]{ 
         \includegraphics[width=5.9cm]{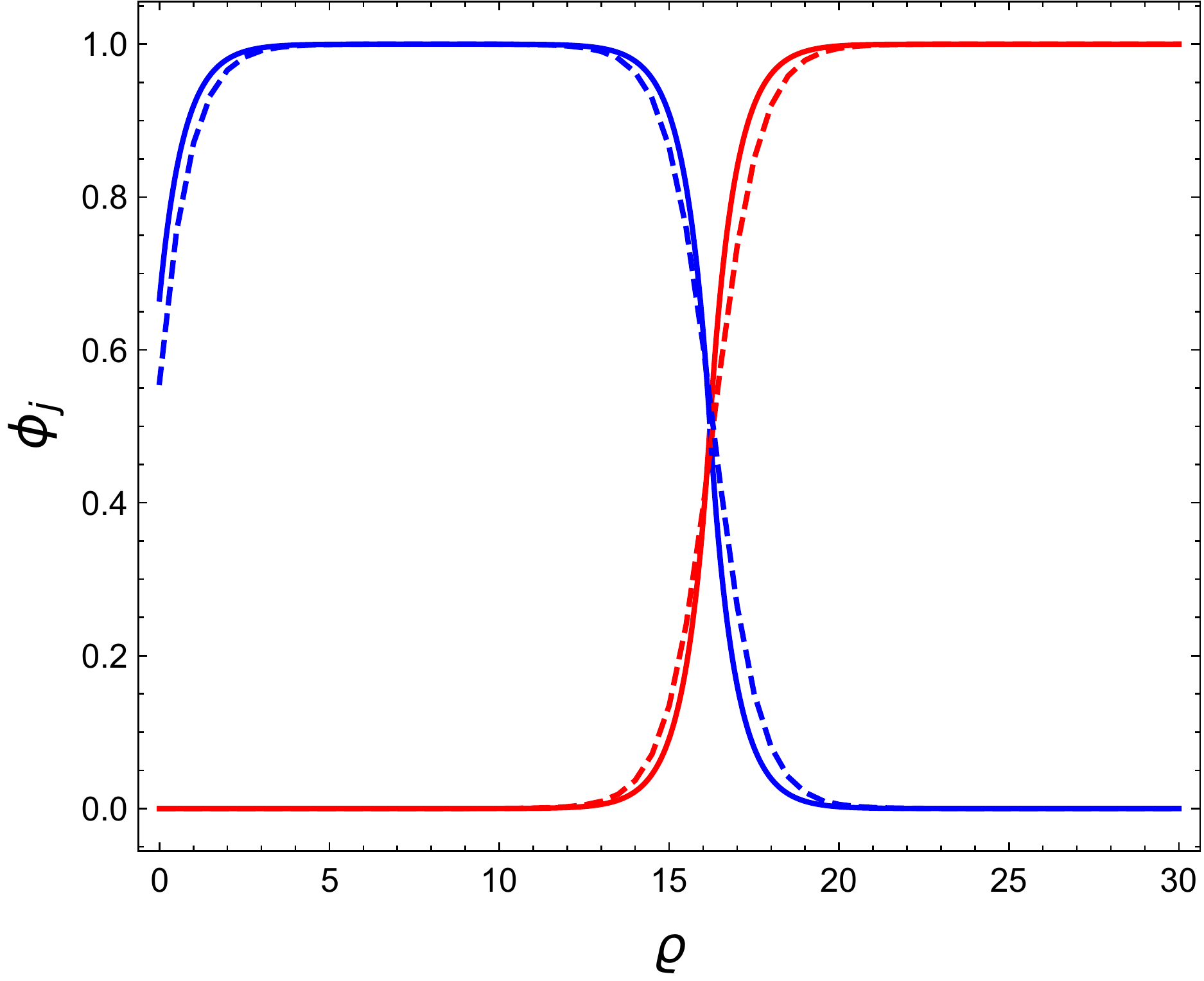} 
        \label{fig:sub-slowed-hopping}}            
        \subfigure[$c = 1/\sqrt 2, h = 0,\xi  = 1.5, K = 3.5.$]{ 
         \includegraphics[width=5.9cm]{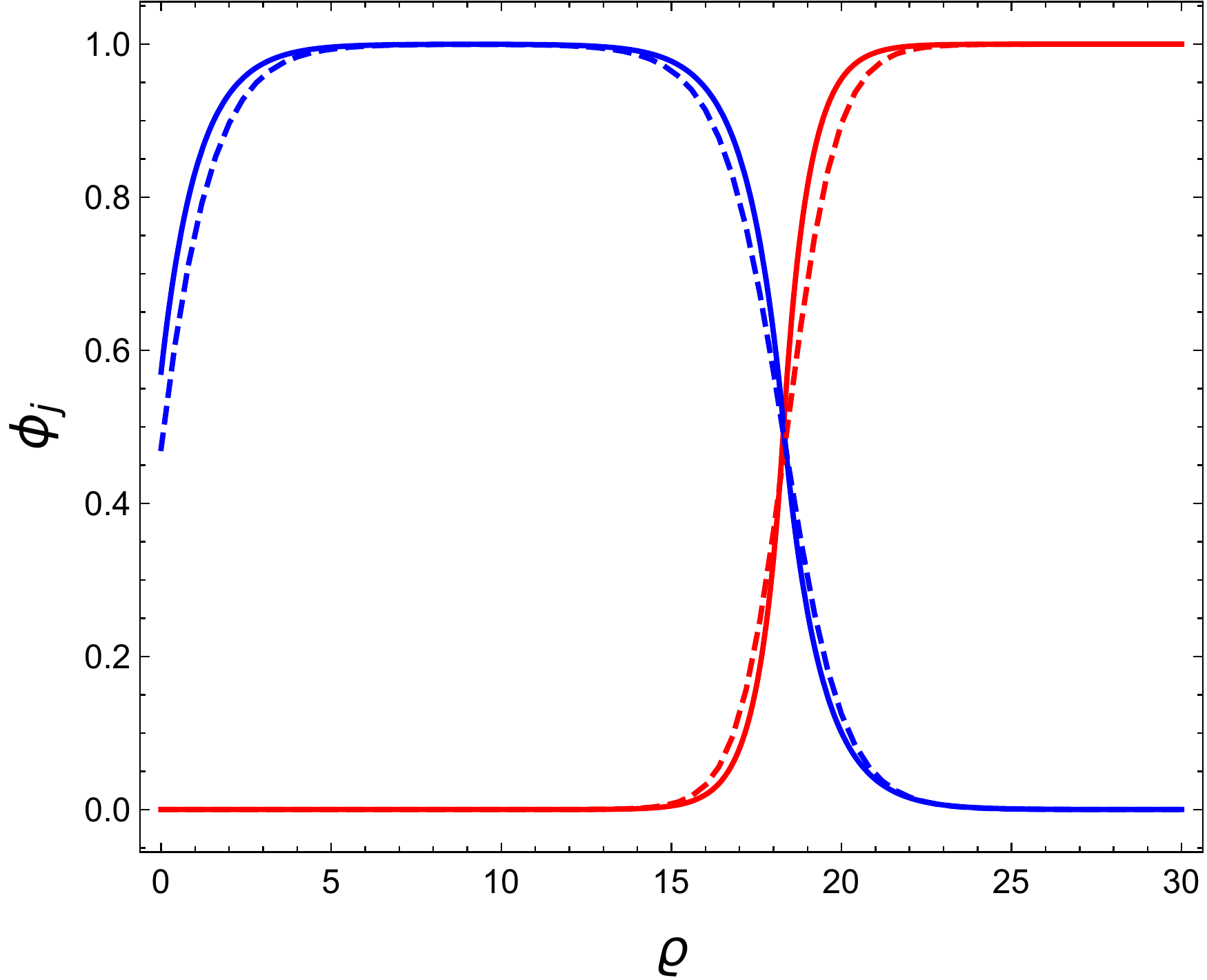}} 
        \label{fig:sub-hybrid-hopping}   
    \caption{(Colour online) The condensate profiles calculated in DPA (solid lines) and in GP equations (dashed lines). The red and blue lines correspond to the first and the second component.} \label{f2}
\end{figure}

The most important requirement for any approximation method is its reliability. To check the reliability of DPA, let us plot in figure~\ref{f2} the condensate profiles corresponding to several values of $\xi $ and at $c = 1/\sqrt 2 $, $h = 0$: a) $\xi  = 1.0$, $K = 3$; b) $\xi  = 0.5$, $K = 2.5$; c) $\xi  = 1.5$, $K = 3.5$. The solid lines denote the solutions (\ref{sl1}), (\ref{sl2}) of  the  DPA equations (\ref{GP3}) and the dashed lines denote the numerical solutions of the GP equations (\ref{GP2}), with the same BCs (\ref{BC4}), (\ref{BC5}) and (\ref{BC6}).

Figure~\ref{f2} tells us that:\\
\indent --	The graphs obtained in DPA are close to those found in the GP theory in the whole variation region of $\varrho$. This implies that the DPA is a reliable method.\\
\indent --	The condensates calculated respectively in DPA and in GP equations provide the same interface position $\ell$.

Note that the foregoing results remain valid for other values of $c$, including $c = 0$ (Neumann BC) and $c = \infty$ (Dirichlet BC). \\
\indent -- Finally, let us calculate the value $c$ appearing in the Robin BC (\ref{BC5}). At first, let us look for ${\phi _2}\left( { - h} \right)$ from (\ref{sl2})
\begin{align}\label{fc1}
{\phi _{2L}}(\varrho  =  - h) = \frac{{1 + {\rm{4}}{{\rm{e}}^{ - \frac{{\sqrt 2 \left( {3h + 2\ell } \right)}}{\xi }}}\left( {M_{1} + M_{2} + M_{3}} \right)}}{{\big( {\sqrt 2  + c\xi } \big)\left\{ {\sqrt 2 \cosh \left[ {\frac{{\sqrt 2 \left( {h + \ell } \right)}}{\xi }} \right] + c\xi \sinh \left[ {\frac{{\sqrt 2 \left( {h + \ell } \right)}}{\xi }} \right]} \right\}M_{4}}}.
\end{align}
The expressions of ${M_j}$, $1 \leqslant j \leqslant 4$ are given in appendix \ref{B}.

When $h$ tends to infinity, (\ref{fc1}) becomes
\begin{align}\label{fc2}
{\lim _{h \to \infty }}{\phi _2}\left( { - h} \right) = {\phi _2}\left( { - \infty } \right) = \frac{1}{{1 + c\xi /2}}\,,
\end{align}
which must coincide with the BC for ${\phi _2}\left( \varrho \right)$ in the infinite space \cite{c12, c14}, ${\phi _2}\left( { - \infty } \right) = 1$. Hence, we get  $c = 0$ and then the Robin BC (\ref{BC5}) turns out to be the Neumann BC.

\section{Interface tension in grand canonical ensemble}\label{3}
The interface tension was determined in grand canonical ensemble (GCE) and canonical ensemble (CE) by \cite{c21,c21a} and \cite{c12}, accordingly. The system can be viewed as having a direct contact with the bulk reservoirs of condensates and we have ${\mu _j} = {g_{jj}}{n_j}$. The interface tension defined in GCE are equal to $4$ times their counterparts in CE \cite{c13n} in the infinite space and different from $4$ times in the semi-infinite space \cite{c15}. To begin with, let us follow closely \cite{c21,c21a} to start from the grand potential of the interface in the dimensionless form calculated at the bulk coexistence of two condensates
\begin{align}
\Omega=2P_0A\int_{-h}^{+\infty}\rd\varrho\Bigg[\sum_{j=1,2}\left(-\phi_j\partial_{\varrho}^2\phi_j\right)+\tilde{{ V}}\Bigg],\label{the5}
\end{align}
where  $P_1=P_2=P_0=g_{jj}n_{j}^2/2$, $A$ is the interface  area and $\tilde V$ is given by (\ref{the3}).

Replacing $\tilde V$ by the DPA potential (\ref{the4}) in equation~(\ref{the5}), we arrive at the  grand potential in DPA
\begin{align}
\Omega=2P_0\xi_1A\int_{- h}^\infty \rd\varrho \big(-\phi_1^*\partial_\varrho^2\phi_1-\xi^2\phi_2^*\partial_\varrho^2\phi_2+\tilde{{ V}}_\text{DPA}\big).\label{the7}
\end{align}
Then, taking into account (\ref{the4}), (\ref{GP3}) and (\ref{the7}), the interface tension is determined as the excess grand potential per unit area, 
\begin{align}
{\gamma_{12}}&=\frac{\Omega-\Omega_\text{b}}{A}=P_0\xi_1\Bigg[\int\limits_{ - h}^\ell  {4(1 - {\phi _{2}})} \rd\varrho  + \int\limits_\ell ^\infty  {4(1 - {\phi _{1{\rm{}}}})} \rd\varrho \Bigg],\label{tens1}
\end{align}
with ${\Omega _\text{b}} = - PV$ being the bulk grand potential.

Substituting (\ref{sl1}), (\ref{sl2}) into (\ref{tens1}), leads to
\begin{align}
{{\tilde \gamma }_{12}} = \frac{{{\gamma _{12}}}}{{{P_0}{\xi _1}}}
=  - 2\sqrt 2 {A_1}{{\rm{e}}^{ - \sqrt 2 \ell }} - \frac{1}{{\sqrt 2  + c\xi }}X,
\label{tens2}
\end{align}
in which
\begin{align*}
{X} &= 2{{\rm{e}}^{ - \frac{{\sqrt 2 \left( {2h + \ell } \right)}}{\xi }}}\left[ { - 1 + {{\rm{e}}^{\frac{{\sqrt 2 \left( {h + \ell } \right)}}{\xi }}}} \right]\xi  - c {{\rm{e}}^{\frac{{\sqrt 2 h}}{\xi }}}\xi+ {B_2}\left[ {\sqrt 2  - c\xi  + {{\rm{e}}^{\frac{{\sqrt 2 \left( {h + \ell } \right)}}{\xi }}}\big( {\sqrt 2  + c\xi } \big)} \right].
\end{align*}
(\ref{tens2}) is the interface tension with Robin BC.

When $c$ tends to infinity, equation~(\ref{tens2}) turns out to be
\begin{align}\label{tens3}
{\tilde \gamma _{12}} = \frac{{2\sqrt 2 \beta }}{{\beta  + \sqrt 2 \tanh \left[ {\left( {h + \ell } \right)\beta } \right]}} + \frac{{4\left[ { - 1 + {{\rm{e}}^{\frac{{\sqrt 2 \left( {h + \ell } \right)}}{\xi }}}} \right] }}{{\sqrt 2  - \beta  + {{\rm{e}}^{\frac{{2\sqrt 2 \left( {h + \ell } \right)}}{\xi }}}\big( {\sqrt 2  + \beta } \big)}} \left[ {1 - \sqrt 2 \beta  + {{\rm{e}}^{\frac{{\sqrt 2 \left( {h + \ell } \right)}}{\xi }}}\big( {1 + \sqrt 2 \beta } \big)} \right]\xi.
\end{align}
Expression (\ref{tens3}) totally coincides with the interface tension calculated imposing the Dirichlet condition at a hard wall \cite{c15}.
When $c$ tends to zero, equation~(\ref{tens2}) reduces to the interface tension with Neumann BC 
\begin{align}\label{tens4}
{{ \tilde\gamma }_{12}}&= {\frac{{4\xi \beta }}{{2 + \sqrt 2 \beta \coth \left[ {\frac{{\sqrt 2 \left( {h + \ell } \right)}}{\xi }} \right]}} + \frac{{2\sqrt 2 \beta }}{{\beta  + \sqrt 2 \tanh \left[ {\left( {h + \ell } \right)\beta } \right]}}}.
\end{align}
In order to better understand the physical significance of the Neumann BC it is convenient to let $h = 0$ and rewrite the interface tension  (\ref{tens1}) in the following form
\begin{align}\label{tens5}
{\tilde \gamma _{12}} &= \frac{{{\gamma _{12}}}}{{{P_0}{\xi _1}}} = 4\int\limits_{ 0}^\ell  {(1 - {\phi _2})\rd\varrho }  + 4\int\limits_\ell ^\infty  {(1 - {\phi _1})\rd\varrho } = b - 4\int\limits_{ 0}^\ell  {{\phi _2}\rd\varrho },
\end{align} 
where
\begin{align*}
b = \frac{{2\sqrt 2 \beta }}{{\beta  + \sqrt 2 \tanh(\beta \ell )}} + 4 \ell >0.
\end{align*}

\begin{figure}[!t]
	\centering
		\includegraphics[scale=0.6]{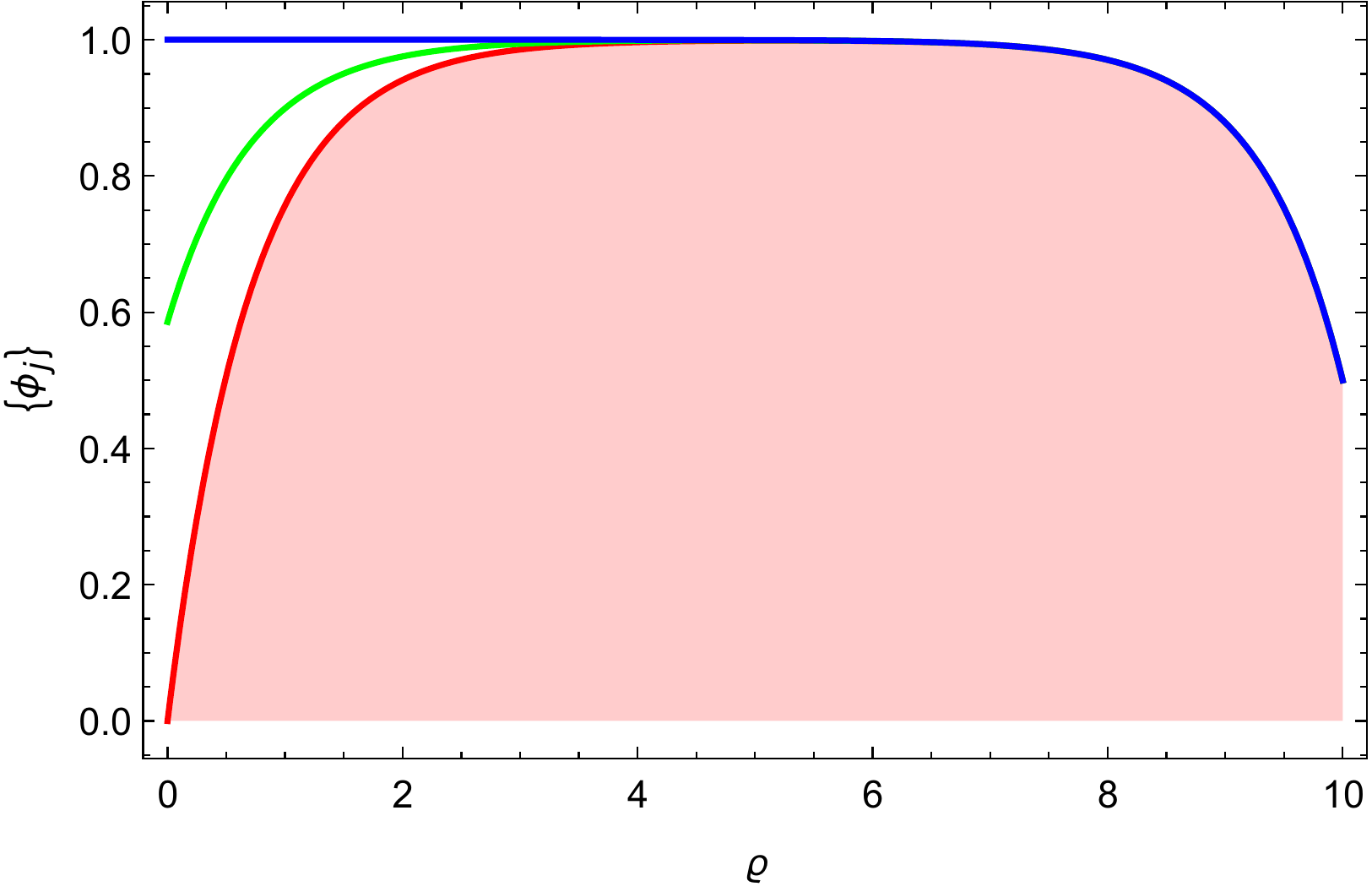}
		\caption{ (Colour online) The profiles of condensate $2$ are plotted in the interval $ 0 \leqslant \varrho  \leqslant \ell   $ at $K = 3$ and $\xi =1$ and several values of $ c $,  $c = 0$ (blue line), $c = 1$ (green line) and $c =\infty$ (red line).}\label{f3}
\end{figure}

The expression $\int\nolimits_{ 0}^\ell {{\phi _2}} \rd\varrho $ is exactly the area of the shaded domain limited by the condensate $2$ and the interval  $0 \leqslant \varrho  \leqslant \ell$ in the $\varrho$ axis, as is shown in figure~\ref{f3}. It is easily realized that the area monotonously decreases for $c$ changing from Neumann BC (blue line), $c = 0$, to Robin BC (green line), $c = 1$, for example, and then to Dirichlet BC (red line), $c\rightarrow\infty$. This implies that the interface tension (\ref{tens5}) obeys the inequality
\begin{align}\label{sosanh1}
{\gamma _{12}}\left( \text{{Neumann}} \right) < {\gamma _{12}}\left(\text {Robin} \right) < {\gamma _{12}}\left(\text{Dirichlet}\right).
\end{align}

\begin{figure}[!t]
	\begin{center}
		\includegraphics[scale=0.5]{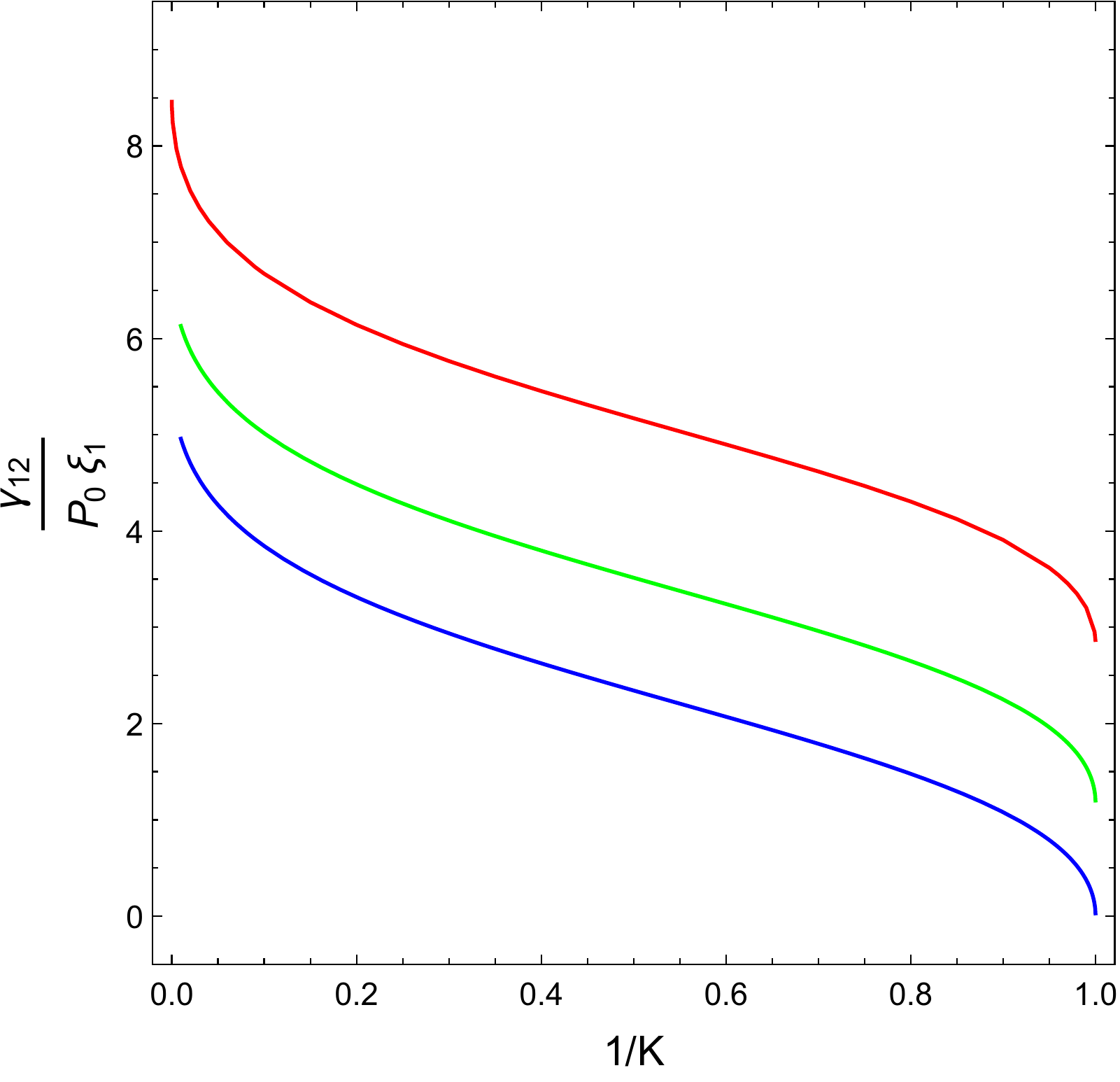}
		\caption{ (Colour online) The evolution of the GCE interface tension versus $K$  at $\xi=1$. The blue, green and red lines correspond, respectively, to Neumann, Robin and Dirichlet BCs.}\label{f4}
	\end{center}
\end{figure}

\begin{figure}[!t]
	\begin{center}
		\includegraphics[scale=0.5]{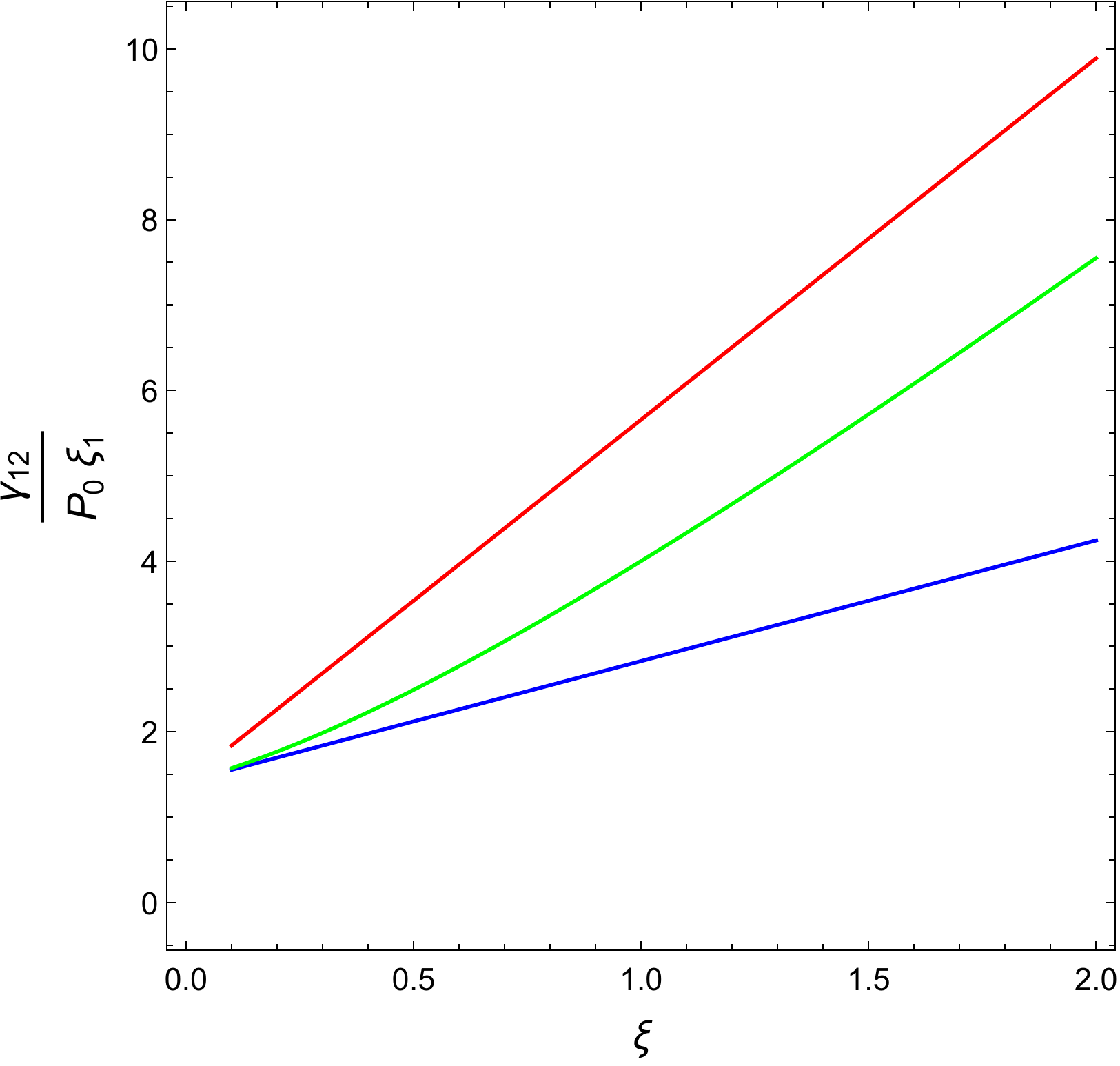}
		\caption{ (Colour online) The evolution of the GCE interface tension versus $\xi$  at $K=3$. The blue, green and red lines correspond, respectively, to Neumann, Robin and Dirichlet BCs.}\label{f42}
	\end{center}
\end{figure}

The above inequality is illustrated in figure~\ref{f4} which displays the $K$ dependence of interface tensions (\ref{tens2}), (\ref{tens3}) and (\ref{tens4}), corresponding,  respectively, to the Robin, Dirichlet and Neumann BCs  at $\xi  = 1$. For other values of  $\xi$, the above inequality (\ref{sosanh1}) remains unchanged (illustrated in figure~\ref{f42}).    

Combining the inequality~(\ref{sosanh1}) and the figure~\ref{f4}, we arrive at the important statement:\\
\indent a)	The state derived from the Neumann BC is  stable while those which are derived from the other BCs are unstable.\\
\indent b)
\begin{align}\label{gh}
{\lim _{K \to 1}}\frac{{\partial {{\tilde \gamma }_{12}}}}{{\partial K}} = \infty. 
\end{align}

\begin{figure}[!t]
	\begin{center}
		\includegraphics[scale=0.8]{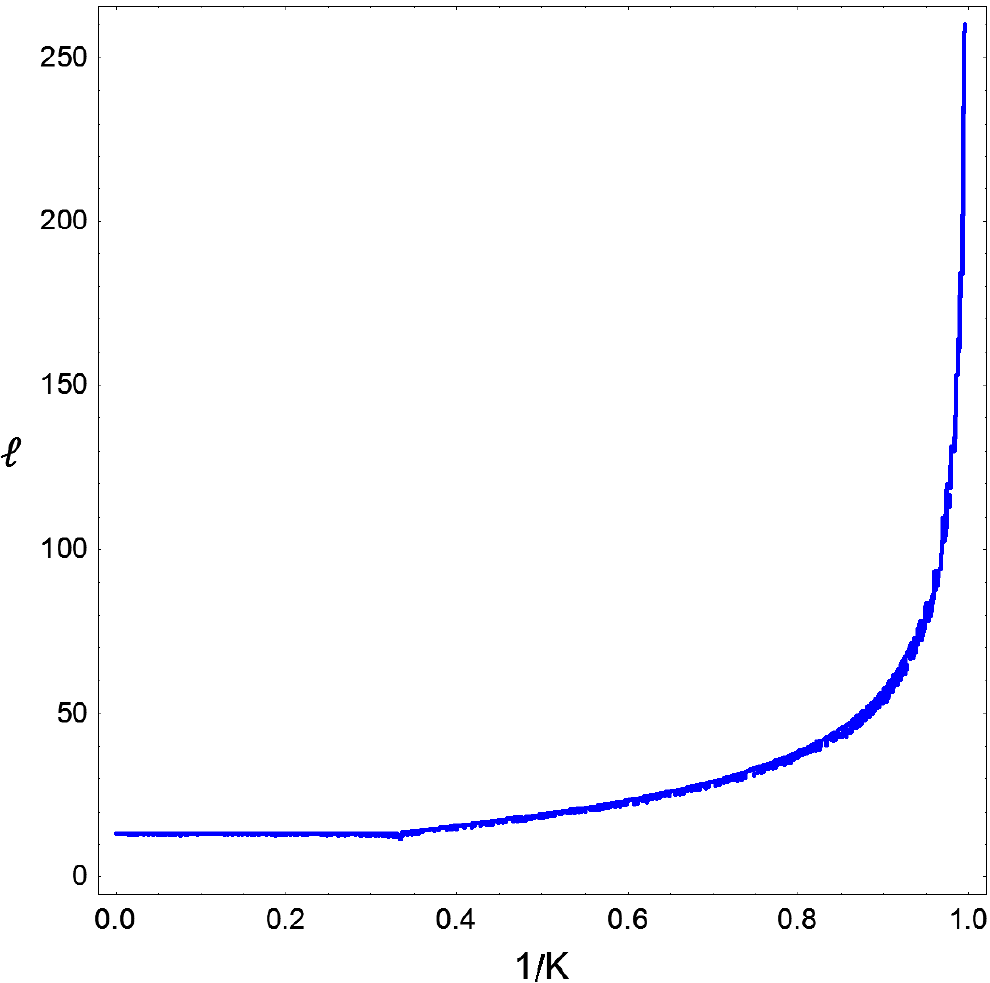}
		\caption{ (Colour online) The evolution of $\ell$ versus $K$  at $\xi = 1.$}\label{f5}
	\end{center}
\end{figure}

Next, let us consider whether or not the system would undergo a transition from immiscible to miscible states when $K$ tends to $1$. To this end, we numerically compute  the evolution of the function $\ell = f(K,\xi )$ given in (\ref{itf2}) versus $K$ and the result is visualized  in figure~\ref{f5}, which gives
\begin{align}\label{ell2}
{\lim _{K \to 1}}\ell\left( {K,\xi  = 1} \right) = \infty .
\end{align}
For other values of $\xi$, the preceding results remain unchanged. Based on (\ref{ell2}), we immediately have   
\begin{align}\label{pha1}
{\lim _{K \to 1}}{\tilde \gamma _{12}} =  {\lim _{K \to 1}}\frac{{2\sqrt 2 }}{{1 + \sqrt 2 \left[ {h + \ell\left( {K,\xi } \right)} \right]}} = 0,
\end{align}
which is valid for all interface tensions (\ref{tens2}), (\ref{tens3}) and (\ref{tens4}). 

Combining (\ref{gh}) and (\ref{pha1}) proves that as $K$ tends to $1$, there occurs a first-order phase transition from immiscible to miscible states.

\begin{figure}[!t]
	\begin{center}
		\includegraphics[width=9cm]{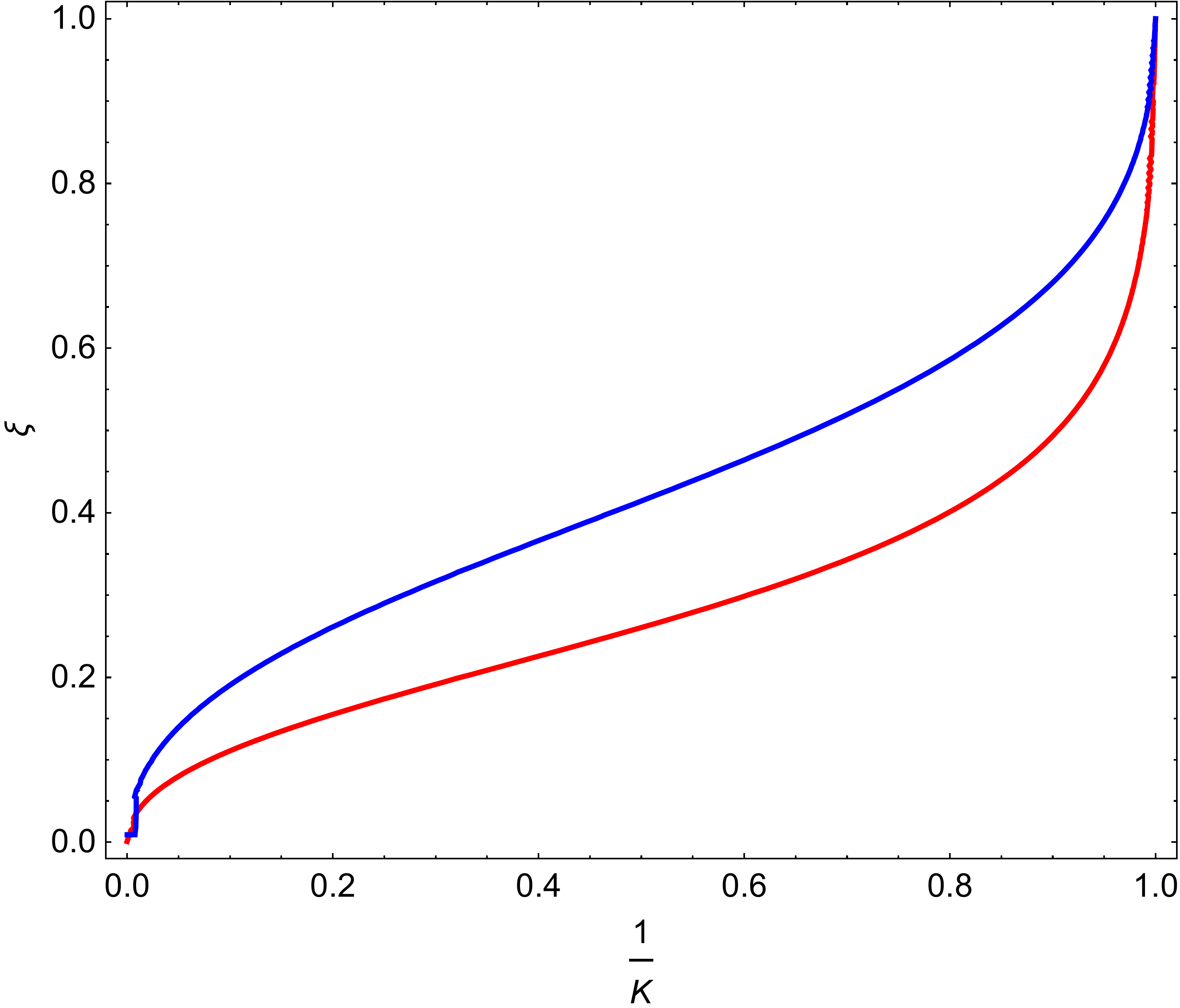}
		\caption{(Colour online) Wetting phase transition at $h = 0$. The red line (blue line) corresponds to the Robin (Dirichlet) BC.}\label{f6}
	\end{center}
\end{figure}

To proceed to the wetting phase transition, let us remark that this phase transition could be realized in the system which was initially in the unstable state. In our case, the wetting phase transition could equally depart from one of the two unstable states as follows: \\
\indent a)	Ground state derived from the Dirichlet BC. We consider the behaviour of the corresponding interface tension given in (\ref{tens3}) in strong segregation
\begin{align}\label{w1}
{\lim _{K \to \infty }}{\gamma _{12}} = 2\sqrt 2P_0\xi_1  + 4\sqrt 2 P_0\xi_2 \tanh \left( {\frac{{h + \ell }}{{\sqrt 2 \xi }}} \right).
\end{align}
On the other hand, for strong segregation, the behaviour ${\gamma _{12}}$ is expressed through the sum of two wall tensions ${\gamma _{\text{W}j}}$ $(j = 1,2)$
\begin{align}\label{w2}
 \mathop {\lim }\limits_{K \to \infty } {\gamma _{12}} = {\gamma _{{\rm{W}}_{1}}} + {\gamma _{{\rm{W_{2}}}}}. 
\end{align}
Comparing (\ref{w1}) and (\ref{w2}), immediately gives
\begin{subequations}
\begin{eqnarray}
&&{\gamma _{{\rm{W}}_{1}}} = 2\sqrt 2 P_0\xi_1\,,\label{w11}\\
&&{\gamma _{{\rm{W_{2}}}}} = 4\sqrt 2 P_0\xi_2 \tanh \left( {\frac{{h + \ell }}{{\sqrt 2 \xi }}} \right).\label{wa22}
\end{eqnarray}
\end{subequations}
The wetting phase transition takes place when the Antonov rule is realized \cite{c13n, c15}
\begin{eqnarray}
\gamma_{\text{W}_{1}}=\gamma_{\text{W}_{2}}+\gamma_{12}.\label{w3}
\end{eqnarray}
The phase diagram of (\ref{w3}) is plotted in figure~\ref{f6} with blue line.

\indent b)	Ground state derived from the Robin BC. Analogously, the strong segregation behaviour of  the interface tension derived from the Robin BC (\ref{tens2}) reads
\begin{align}\label{w4}
{\lim _{K \to \infty }}{ \gamma _{12}} &= 2\sqrt 2 P_0\xi_1\nonumber\\
&+4{P_0}{\xi _2}\frac{{2c{{\rm{e}}^{\frac{{2\sqrt 2 \ell }}{\xi }}}\xi ( {\sqrt 2  + c\xi } ) + 2\sqrt 2 c{{\rm{e}}^{\frac{{\sqrt 2 \ell }}{\xi }}}\xi ( { - 2 + {c^2}{\xi ^2}} ) - ( { - 1 + \sqrt 2 c\xi } )( { - 2 + {c^2}{\xi ^2}} ) - {Y}}}{{( {\sqrt 2  + c\xi } ){{\left[ {\sqrt 2  - c\xi  + {{\rm{e}}^{\frac{{2\sqrt 2 \ell }}{\xi }}}( {\sqrt 2  + c\xi } )} \right]}^2}}}
\end{align}
with
\begin{align*}
Y =  - 2c{{\rm{e}}^{\frac{{3\sqrt 2 \ell }}{\xi }}}\xi \left[ {2\sqrt 2  + c\xi ( {4 + \sqrt 2 c\xi } )} \right] + {{\rm{e}}^{\frac{{4\sqrt 2 \ell }}{\xi }}}\left\{ {2 + c\xi \left[ {4\sqrt 2  + c\xi ( {5 + \sqrt 2 c\xi } )} \right]} \right\}.
\end{align*}
Combining (\ref{w2}) and (\ref{w4}), yields
\begin{subequations}
\begin{align}
&{\gamma _{{\rm{w}}_{1}}} = 2\sqrt 2 P_0\xi_1\,,\label{wall1}\\
&{\gamma _{{\rm{w_{2}}}}} = 4{P_0}{\xi _2}\frac{{2c{{\rm{e}}^{\frac{{2\sqrt 2 \ell }}{\xi }}}\xi ( {\sqrt 2  + c\xi } ) + 2\sqrt 2 c{{\rm{e}}^{\frac{{\sqrt 2 \ell }}{\xi }}}\xi ( { - 2 + {c^2}{\xi ^2}} ) - ( { - 1 + \sqrt 2 c\xi } )( { - 2 + {c^2}{\xi ^2}} ) - {Y}}}{{( {\sqrt 2  + c\xi } ){{\left[ {\sqrt 2  - c\xi  + {{\rm{e}}^{\frac{{2\sqrt 2 \ell }}{\xi }}}( {\sqrt 2  + c\xi } )} \right]}^2}}}.\label{wall2}
\end{align}
\end{subequations}
The phase diagram of the wetting phase transition determined by the Antonov rule (\ref{w3}) is depicted in figure~\ref{f6} by red line. There is easily seen a distinction between two diagrams corresponding to two different  phase transitions. Although both phase transitions are possible, but the one associated with the Robin BC is more favourable because it corresponds to the smaller interface tension (\ref{sosanh1}).

\section{Conclusion and outlook}\label{4}
Based on  the GP equations in the DPA, we investigated the system of two segregated Bose-Einstein condensates restricted by a hard wall. Our main results are in order:\\
\indent a)	Making use of the consistency between the the BCs imposed on the condensates in confined geometry and in infinite space, we determined the BCs for both condensates at a hard wall, namely, the Neumann BC for the second condensate and Dirichlet BC for the first condensate.\\
\indent b)	The analytical expressions for both condensate profiles, calculated in different BCs, are very close to those derived from the GP equations by numerical computations. This proves the reliability of the DPA.\\
\indent c)	Based on the analytical expressions of interface tension, calculated in different BCs, we showed that $${\gamma _{12}}\left( \text{{Neumann}} \right) < {\gamma _{12}}\left(\text {Robin} \right) < {\gamma _{12}}\left(\text{Dirichlet}\right),$$ which proves that the ground state corresponds to the Neumann BC and is stable while other two BCs lead to unstable states.\\
\indent d) The above mentioned result leads to an important consequence: In our system there exist two scenarios of wetting phase transitions  which correspond to two different unstable states, but the one associated with the Robin unstable state is more favourable because it corresponds to a smaller interface tension. \\
\indent e)	As $K$ tends to $1$, the system undergoes a first-order phase transition from immiscible to miscible states.

The results mentioned in items a), c) and d) are our major successes, they are very meaningful because the system of binary BECs mixture is also useful for many other systems of great interest, such as multi-component superconductors, Mott-insulators and so on \cite{c24}. Especially, it is very interesting to extend the method presented in this work, in a straightforward manner, to two-gap superconductors \cite{c25} whose Hamiltonians have the same structure of our system. It is expected that the application of the DPA to these systems could provide new interesting effects. 

Last but not least, it is worth to remark that our previous results were derived within the framework of the Gross-Pitaevskii theory based on the mean field approximation, whereby the thermal and quantum fluctuations were completely neglected \cite{c26, c27}. Although a great number of the effects related to the Bose-Einstein condensations can be successfully explained by the Gross-Pitaevskii theory, in recent years a more complete theory is most welcome in order to comprehend various experiments associated with quantum fluctuations, solitons, vortices and so on \cite{c28, c29, c30}.

\section*{Acknowledgements}
This paper is supported by the Vietnam Ministry of Education and Training in the framework of the Scientific Research Project  under Grant B 2015-25-33. The discussion with Pham The Song is acknowledged with thanks.

\appendix
\section{The analytical expressions of $A_j$, $B_j$ $(j =1, 2)$}\label{A}
\begin{align*}
A_1&= - \frac{{{{\rm{e}}^{\sqrt 2 \ell }}\beta }}{{\beta  + \sqrt 2 \tanh \left[ {\left( {h + \ell } \right)\beta } \right]}}\,, \quad B_1= \frac{{\sqrt 2 {{\rm{e}}^{\left( {2h + \ell } \right)\beta }}}}{{ - \sqrt 2  + \beta  + {{\rm{e}}^{2\left( {h + \ell } \right)\beta }}\big( {\sqrt 2  + \beta } \big)}}\,,\\
A_2&= \frac{{{{\rm{e}}^{\frac{{\ell \beta }}{\xi }}}\Big[ { - 1 + {{\rm{e}}^{\frac{{\sqrt 2 \left( {h + \ell } \right)}}{\xi }}}} \Big]{A_{21}}}}{{\big( {\sqrt 2  + c\xi } \big)\big( { - 2 + \sqrt 2 \beta  + {A_{22}}} \big)}}\,, \quad B_2=\frac{{{B_{21}} + {{\rm{e}}^{\frac{{\sqrt 2 \left( {2h + \ell } \right)}}{\xi }}}\beta \left( { - 2 + {c^2}{\xi ^2}} \right) - {B_{22}} - {B_{23}}}}{{{B_{24}}\big[ { - 2 + \sqrt 2 \beta  + c\big( {\sqrt 2  - \beta } \big)\xi  + {B_{25}}} \big]}}\,,
\\
{A_{21}} &= \sqrt 2 \big( {2 - {c^2}{\xi ^2}} \big) + {{\rm{e}}^{\frac{{\sqrt 2 ( {h + \ell } )}}{\xi }}}\big[ {2\sqrt 2  + c\xi \big( {4 + \sqrt 2 c\xi } \big)} \big], 
\end{align*}
\begin{align*}
{A_{22}} &= c\big( {\sqrt 2  - \beta } \big)\xi  + {{\rm{e}}^{\frac{{2\sqrt 2 ( {h + \ell } )}}{\xi }}}\big[ {2 + \sqrt 2 \beta  + c\big( {\sqrt 2  + \beta } \big)\xi } \big], \\
{B_{21}}&= c{{\rm{e}}^{\frac{{\sqrt 2 h}}{\xi }}}\xi \big[ { - 2 + \sqrt 2 \beta  + c\big( {\sqrt 2  - \beta } \big)\xi } \big], \\
{B_{22}} &= c{{\rm{e}}^{\frac{{\sqrt 2 ( {3h + 2\ell } )}}{\xi }}}\xi \big[ {2 + \sqrt 2 c\xi  - \beta \big( {\sqrt 2  + c\xi } \big)} \big], \quad {B_{23}} = {{\rm{e}}^{\frac{{\sqrt 2 ( {4h + 3\ell } )}}{\xi }}}\beta \big[ {2 + c\xi \big( {2\sqrt 2  + c\xi } \big)} \big], \\
{B_{24}} &= \sqrt 2  - c\xi  + {{\rm{e}}^{\frac{{2\sqrt 2 ( {h + \ell } )}}{\xi }}}\big( {\sqrt 2  + c\xi } \big), \quad {B_{25}} = {{\rm{e}}^{\frac{{2\sqrt 2 ( {h + \ell } )}}{\xi }}}\big[ {2 + \sqrt 2 \beta  + c\big( {\sqrt 2  + \beta } \big)\xi } \big].
\end{align*}
\section{ The analytical expressions of ${M_j}$, $1 \leqslant j \leqslant 4 $}\label{B}
\begin{align*}
{M_1} &= 2c{{\rm{e}}^{\frac{{\sqrt 2 \left( {3h + 2\ell } \right)}}{\xi }}}\xi \big( {\sqrt 2  + c\xi } \big)\left( { - 2 + c\beta \xi } \right) + 2\sqrt 2 {{\rm{e}}^{\frac{{\sqrt 2 \left( {2h + \ell } \right)}}{\xi }}}\beta \big( { - 2 + {c^2}{\xi ^2}} \big),\ \\
{M_2} &=  - c{{\rm{e}}^{\frac{{\sqrt 2 \left( {5h + 4\ell } \right)}}{\xi }}}\xi \big[ {2\big( {\sqrt 2  + \beta } \big) + 2c\big( {2 + \sqrt 2 \beta } \big)\xi  + {c^2}\big( {\sqrt 2  + \beta } \big){\xi ^2}} \big],\ 
\end{align*}
\begin{align*}
{M_3}& = c{{\rm{e}}^{\frac{{\sqrt 2 h}}{\xi }}}\big( {\sqrt 2  - \beta } \big)\xi \big( { - 2 + {c^2}{\xi ^2}} \big) - 2{{\rm{e}}^{\frac{{\sqrt 2 \left( {4h + 3\ell } \right)}}{\xi }}}\beta \big[ {2\sqrt 2  + c\xi \big( {4 + \sqrt 2 c\xi } \big)} \big],\ \\
{M_4}& = {\sqrt 2 \left( {\beta  + c\xi } \right)\cosh \left[ {\frac{{\sqrt 2 \left( {h + \ell } \right)}}{\xi }} \right] + \left( {2 + c\beta \xi } \right)\sinh \left[ {\frac{{\sqrt 2 \left( {h + \ell } \right)}}{\xi }} \right]} .
\end{align*}

\ukrainianpart

\title{Ефекти скінченного розміру у двох відокремлених  конденсатах Бозе-Ейнштейна, обмежених твердою стінкою}
\author{Х.В. Квует\refaddr{label1}, Н.В. Тху\refaddr{label1}, Д.Т. Там\refaddr{label2}, Т.Х. Пхат\refaddr{label3}}
\addresses{
\addr{label1}Фізичний факультет, Ханойський педагогічний університет 2, Ханой, В'єтнам 
\addr{label2} Університет Тау Бак, Шон Ла, В'єтнам 
\addr{label3} Комісія атомної енергетики  В'єтнаму,  Ханой, В'єтнам 
}

\makeukrtitle

\begin{abstract}
Ефекти скінченного розміру у двох відокремлених  конденсатах Бозе-Ейнштейна, обмежених твердою стінкою, досліджено з допомогою рівнянь Гросса-Пітаєвского в наближенні подвійної параболи.  Виходячи з узгодженості між граничними умовами, які накладаються на конденсати в обмеженій геометрії і в повному просторі, ми знаходимо усі можливі  граничні умови разом із відповідними профілями конденсату і міжфазові натяги.
Нами відкрито два ефекти скінченного розміру:  a) основний стан, отриманий з граничної умови Ньюмана, є стійким, а основні стани, отримані з граничних умов Робіна і Діріхлє,	є нестійкими, b) 
отже, однаковим чином проявляються два можливих переходи змочування як результат двох нестійких станів. 
Проте, перехід, пов'язаний з граничними умовами Робіна, виявляється більш сприятливим, тому що він відповідає меншому міжфазовому натягу.  
\keywords ефекти скінченного розміру, конденсати Бозе-Ейнштейна, гранична умова, наближення подвійної параболи
\end{abstract}
\end{document}